\documentclass[twoside,11pt]{article}

\usepackage{obs_study_style}
\usepackage{amssymb}
\usepackage{amsmath}
\usepackage{booktabs}
\usepackage{multirow}
\usepackage{array}
\usepackage{hyperref}

\heading{}{}{}{}{}{Yijun Fan and Dylan S. Small}

\ShortHeadings{A subgroup-aware scoring approach to the study of effect modification in observational studies}{Fan and Small}
\begin{document}
\firstpageno{1}
\title{A subgroup-aware scoring approach to the study of effect modification in observational studies}

\author{\name Yijun Fan \email yjfan@uchicago.edu \\
       \addr Department of Statistics\\
       University of Chicago\\
       Chicago, IL 60637, USA
       \AND
       \name Dylan S. Small \email dsmall@wharton.upenn.edu \\
       \addr Department of Statistics and Data science\\
       University of Pennsylvania \\
       Philadelphia, PA 19104, USA}

\maketitle

\begin{abstract}
Effect modification means the size of a treatment effect varies with an observed covariate. Generally speaking, a larger treatment effect with more stable error terms is less sensitive to bias. Thus, we might be able to conclude that a study is less sensitive to unmeasured bias by using these subgroups experiencing larger treatment effects. \citet{submax} proposed the submax method that leverages the joint distribution of test statistics from subgroups to draw a firmer conclusion if effect modification occurs. However, one version of the submax method uses M-statistics as the test statistics and is implemented in the \verb|R| package \verb|submax| \citep{submaxpackage}. The scaling factor in the M-statistics is computed using all observations combined across subgroups.  We show that this combining can confuse effect modification with outliers. We propose a novel group M-statistic that scores the matched pairs in each subgroup to tackle the issue. We examine our novel scoring strategy by simulating power in finite samples and by deriving the design sensitivity. The proposed method is applied to an observational study of the effect of a malaria prevention treatment in West Africa.
\end{abstract}


\section{Introduction}
\subsection{Effect modification and sensitivity analysis}
In an observational study, subjects are not randomly assigned to treatment or control, so the treated and control groups might differ in terms of both measured and unmeasured covariates. To make subjects in the treated group and control group comparable, adjustments (e.g. matching) are used to address the concern of measured covariates, but there is still typically concern about unmeasured covariates. Then, a sensitivity analysis asks how large the magnitude of bias from some unmeasured covariates would need to be to explain away the qualitative conclusion of a study based on the assumption of no unmeasured confounding \citep{obs_study}. 

Effect modification occurs when the size of a treatment effect varies depending on a measured covariate. In general, larger treatment effects are less sensitive to unmeasured confounding, which suggests that effect modification can play a role in reducing sensitivity to unmeasured covariates \citep{effect_mod}.  If some subgroups experience larger effects and we make use of those subgroups appropriately, then we may be able to report less sensitivity to unmeasured confounding. The submax method, proposed by \citet{submax}, splits the population into certain overlapped subgroups and uses the joint distribution of these correlated test statistics from the subgroups to study the effect modification in observational studies. For example, a researcher may want to make a robust inference for subgroup analyses by using the submax method based on Huber M-statistics following the suggestion of \citet{mstat}. However, although the submax method considers the subgroups, the scores entering the chosen test statistic (see section \ref{notation} and \ref{normalm}) used by the submax method do not account for such subgroup structures. In this article, we would show that such practice could substantially deteriorate the sensitivity analysis in some cases because effect modification is confused with outliers. We illustrate this point by using M-statistics \citep{mstat} and propose a subgroup-aware scoring approach. 

The motivating example of malaria control in West Africa is introduced in section \ref{garkibg}. The notation and the submax method is reviewed in section \ref{setup}. We introduce the novel scoring approach in section \ref{scoring} and apply the proposed method to the malaria control example in section \ref{garki}. A simulation is provided in section \ref{sim} to evaluate this approach in extensive settings.

\subsection{Motivating example: control of malaria in West Africa}
\label{garkibg}
The World Health Organization, with the Nigerian government, worked on comparing the effectiveness of different treatment methods of controlling malaria in West Africa \citep{garkiintro}. The treatment of interest in our study is spraying with propoxur, an insecticide, and mass administration of a drug called sulfalene-pyrimethamine. The outcome measurement documents the frequency of Plasmodium falciparum, a protozoan parasite that causes malaria, in blood samples. A series of blood samples were collected across a period of time and we used the four surveys immediately before the treatment and four surveys immediately after the treatment to compute a score. Specifically, any individual included was required to have at least two surveys before and after treatment; then these surveys were combined by using Huber's M-estimate as a trimmed mean for pretreatment and posttreatment summaries. We are interested in whether such treatment would cause lower Plasmodium falciparum frequency. The assignment of treatment was based on operational concerns, such as the feasibility of the treatment, geography, logistics, etc \citep{garkiintro}. Only individual-level covariates, namely age and gender, were included in matching. Prior to using the outcome information other than quality control, we did pair matching for age and gender, as in \citet{effect_mod}, which resulted in $I = 1560$ matched pairs.

\section{Setup of sensitivity analysis for subgroup comparisons}
\label{setup}
\subsection{Notation}
\label{notation}
Suppose we are interested in $L$ binary covariates as potential effect modifiers and thus $G = 2^L$ non-overlapping interaction subgroups are defined. 
Specifically, in our example for malaria control, we are interested in $L = 2$ covariates (i.e. age and gender), and consequently $G = 2^2 = 4$ independent interaction groups are defined. 
Consider here we have $I_g$ matched sets in each group $g$, $1\leq g \leq G$, and in each matched set $gi, 1\leq i \leq I_g$, there is one treated subject $j$ and $n_{gi} - 1$ non-treated subjects.
For simplicity, we only consider the case of matched pairs here, that is $n_{gi} = 2$ and we have the measured covariates $x_{gi1} = x_{gi2}$. In addition to the measured covariates, there may be an unmeasured covariate $u_{gi1}\neq u_{gi2}$. Denote $Z_{gij}$ as an indicator of treatment assignment, so we have  $1=Z_{gi1} + Z_{gi2}$. We denote $r_{Tgij}$ and $r_{Cgij}$ as the responses of $j^{th}$ individual in the $i^{th}$ matched pair of group $g$. Following the potential outcome framework, we have the exhibited response for individual $gij$, that is $R_{g i j}=Z_{g i j} r_{Tg i j}+(1-Z_{g i j}) r_{C g i j}$. Write $\mathcal{F} = \{(r_{Tgij}, r_{Cgij}, x_{gij}, u_{gij}), g = 1, \dots, G, i = 1, \dots, I_g, j = 1, 2\}$. Denote $\mathcal{Z}$ as the set containing $|\mathcal{Z}|=\prod_{g=1}^G 2^{I_g}$ possible values $\mathbf{z}$ of all the possible treatment assignments $\mathbf{Z}=(Z_{111}, Z_{112}, \ldots, Z_{G, I_G, 1},Z_{G,I_G,2})^T$, so $\mathbf{z} \in \mathcal{Z}$ if $z_{g i j} \in \{0,1\}$ and $1=z_{g i 1}+z_{gi2}$ for each $g$ and $i$. Conditioning on the event $\mathbf{Z} \in \mathcal{Z}$ is abbreviated as conditioning on $\mathcal{Z}$, and we have $\operatorname{Pr}(\mathbf{Z}=\mathbf{z} \mid \mathcal{F}, \mathcal{Z})=|\mathcal{Z}|^{-1}$ for each $\mathbf{z} \in \mathcal{Z}$. 

We are interested in testing Fisher's sharp null hypothesis of no treatment effect $H_0: r_{Tgij} = r_{Cgij}, g=1,\cdots,G, i=1,\cdots,I_g, j=1,2$. Under the null, we have the observed treated-minus-control difference $D_{gi} = (Z_{gi1} - Z_{gi2})(R_{gi1} - R_{gi2}) = (Z_{gi1} - Z_{gi2})(r_{Cgi1}-r_{Cgi2}) = \pm(r_{Cgi1}-r_{Cgi2}) =  \epsilon_{gi}$ of matched pair $gi$. If treatments were assigned randomly, the randomization creates the exact null distribution of the statistic $T=\sum_{g=1}^G T_g =\sum_{g=1}^G\sum_{i=1}^{I_g} \text{sign}(D_{gi})q_{gi}$ with some properly chosen score function $q_{gi}$, since now the only randomness is from the treatment assignment $\mathbf{Z}$ as we condition on $\mathcal{F}$ and $\mathcal{Z}$. If we fix a subgroup $g$, we can also use the score statistic $T_g = \sum_{i=1}^{I_g} \text{sign}(D_{gi})q_{gi}$ for this subgroup. Specifically, under the null hypothesis, $T_g$ is the summation of $I_g$ independent random variables taking values $q_{gi}$ and $-q_{gi}$ each with probability $1/2$, and the probability $\operatorname{Pr}(T_g \geq t \mid \mathcal{F}, \mathcal{Z})$ is just the proportion of treatment assignment $\mathbf{Z} \in \mathcal{Z}$ with $T_g\geq t$. The score $q_{gi}$ may depend on $|D_{gi}|$. For example, it replicates Maritz's version of Huber M-statistic \citep{mstat} if we use the trimming version of the score (see section \ref{normalm}) and permutational t-test if we use the raw data. 


However, in an observational study, treatment and control are not assigned at random. Formally, in Rosenbaum's sensitivity model \citep{sensitivity,obs_study}, a sensitivity parameter $\Gamma \geq 1$ is introduced. Denote $\pi_{gi j}=\operatorname{Pr}(Z_{gi j}=1 \mid \mathcal{F})$, and we assume
\begin{equation}
\label{eq:1}
\frac{1}{\Gamma} \leq \frac{\pi_{gi j}(1-\pi_{g^{\prime}i^{\prime} j^{\prime}})}{\pi_{g^{\prime}i^{\prime} j^{\prime}}(1-\pi_{gi j})} \leq \Gamma, \text{ whenever } \mathbf{x}_{gi j}=\mathbf{x}_{g^{\prime}i^{\prime} j^{\prime}};
\end{equation}
$\Gamma$ is the maximum odds ratio of getting treatment for two units with the same measured covariates but a different unmeasured covariate. $\Gamma =1$ is a randomized experiment; larger $\Gamma$'s allow for more unmeasured confounding. Given a $\Gamma$, we obtain a range of possible significance levels (i.e. $P$-values) under this constraint on the treatment assignment. Based on the asymptotic separability technique \citep{sep_sen}, we could obtain an approximation to the upper bound for $P$-values. Formally, denote $\mu_{\Gamma g}$ as the maximum expectation of $T_g$ and $\nu_{\Gamma g}$ as the maximum variance of $T_g$ that achieves the maximum expectation $\mu_{\Gamma g}$ under this constraint. When $\min(I_g) \to \infty$ together with some regularity conditions, we have the following approximated $P-$value upper bound
\begin{equation}
\label{eq:2}
1-\Phi\left\{\left(\sum_{g=1}^G T_g-\mu_{\Gamma g}\right) / \sqrt{\sum_{g=1}^G \nu_{\Gamma g}}\right\},
\end{equation}
where $\Phi$ is the cumulative distribution function of the standard normal distribution. Moreover, now that we have $\min(I_g) \to \infty$ and $G$ groups are independent, it holds that the joint distribution of $G$ statistics $(T_g - \mu_{\Gamma g})/\nu_{\Gamma g}^{1/2}$ is a $G-$dimensional standard normal distribution.

\subsection{The submax method: joint bounds of subgroup comparisons}
\label{submax}
Now that we wish to jointly evaluate multiple subgroups, the submax method \citep{submax} uses the maximum standard deviates of multiple subgroups as the test statistic, whose distribution under the null could be derived explicitly in the large sample as $\min(I_g) \to \infty$ by manipulating the $G-$dimensional standard normal distribution mentioned above. However, instead of conducting $G = 2^L$ independent tests for these finest groups, the submax method only conducts $2L$ subgroup tests plus one overall test, resulting in $K = 2L+1$ correlated tests in total. In the running example, the submax method would conduct $K = 2\times 2+1 = 5$ correlated tests for all the subjects, subjects with age less than $10$, subjects with age larger than $10$, female, and male respectively. Formally, suppose we are interested in making $K = 2L+1$ comparisons, and we define the $K \times G$ matrix $\mathbf{C}$ such that each row $\mathbf{c}_{k*} = (c_{k1},\ldots,c_{kG})^{T}, 1\leq k \leq K$ encodes whether a group $g, 1\leq g \leq G, $ is included in the $k^{th}$ comparison. Then, for this comparison, we have the score $S_k = \sum_{g=1}^G c_{kg}T_g$. Write $\boldsymbol{\mu}_{\Gamma} = (\mu_{\Gamma 1},\ldots,\mu_{\Gamma_G})^T$ and $\mathbf{V}_{\Gamma} = \text{diag}\{\nu_{\Gamma 1}\ldots \nu_{\Gamma G }\}$. Thus, we have the joint distribution of $(S_1,\ldots , S_K)$ as $N(\mathbf{C} \boldsymbol{\mu}_{\Gamma},\mathbf{C V}_{\Gamma} \mathbf{C}^T)$. Denote $\boldsymbol{\theta}_{\Gamma}=\mathbf{C} \boldsymbol{\mu}_{\Gamma}$ and $\boldsymbol{\Sigma}_{\Gamma}=$ $\mathbf{C V}_{\Gamma} \mathbf{C}^T$ while we write $\theta_{\Gamma k}$ as the $k^{th}$ coordinate of $\boldsymbol{\theta}_{\Gamma}$ and $\sigma_{\Gamma k}$ as the $k^{th}$ diagonal element of $\boldsymbol{\Sigma}_{\Gamma}$. Finally, denote $D_{\Gamma k}=\left(S_k-\theta_{\Gamma k}\right) / \sigma_{\Gamma k}$, and we have the joint distribution of $\mathbf{D}_{\Gamma}=\left(D_{\Gamma 1}, \ldots, D_{\Gamma K}\right)^T$ as $N(0,\boldsymbol{\rho}_{\Gamma})$, where $\boldsymbol{\rho}_{\Gamma}$ is the $K\times K$ correlation matrix whose element in $i^{th}$ row and $j^{th}$ column is derived from dividing $(\boldsymbol{\Sigma}_{\Gamma})_{ij}$ by $\sigma_{\Gamma i}$ and $\sigma_{\Gamma j}$. We now have the null distribution of the test statistic $D_{\Gamma \max }=\max _{1 \leq k \leq K} D_{\Gamma k}$ used in the submax method. The critical value $\kappa_{\Gamma, \alpha}$ at level $\alpha$  for $D_{\Gamma \max }$ solves $1-\alpha =\operatorname{Pr}\left(D_{\Gamma \max }<\kappa_{\Gamma, \alpha}\right) $. In particular, in the case of matched pairs, it holds that $\boldsymbol{\rho}_{\Gamma} = \boldsymbol{\rho}$ would not depend on $\Gamma$ (it only depends on the observed data). Thus, given the input scores, the corresponding critical value $\kappa_{\Gamma, \alpha}=\kappa_{\alpha}$ does not vary with $\Gamma$ in the case of matched pairs.

\section{Subgroup-aware scoring approach}
\label{scoring}
\subsection{M-score by trimming together}
\label{normalm}
Before introducing our proposed method, it might be helpful to review the conventional practice when using Maritz-Huber's M-statistic in the submax method. Suppose we are interested in testing the Fisher's sharp null hypothesis of no treatment effect $H_{0}$ using M-statistic, so we plug in the M-score. Denote $h_0 = \text{median}\{|D_{11}|,\cdots,|D_{1I_1}|,\cdots, |D_{gIg}|\}$, which is the median of the absolute values of all $I$ treated-minus-control differences, and we have the score test statistic for group $g, 1\leq g \leq G$,
\begin{equation}
\label{eq:3}
T_g= \sum_{i=1}^{I_g} sign(D_{gi})\psi(\frac{|D_{gi}|}{h_{0}}),
\end{equation}
where $\psi(\cdot)$ is an odd function and $\psi(d)$ is non-negative when $d>0$; $sign(\cdot)$ indicates the sign. Common in the practice is the trimming version of $\psi(\cdot)$ suggested by \citet{mstat} and further discussion about its application in sensitivity analysis is provided in \citet{sensm}. Specifically, $\psi(d) = 0$ when $|d|<a$ while $\psi(d) = d$ when $a<|d|\leq t$ and $\psi(d) = t$ when $|d|>t$. $a$ is the so-called inner parameter \citep{impact_multiple_m} and $t$ is the so-called trimming parameter. In this article, we follow the default setting of function \verb|Mscorev| in the \verb|R| package \verb|submax| \citep{submaxpackage} where $a = 0$ and $t = 3$ when carrying out the experiments. The submax method would then use the $G$ score statistics in equation (\ref{eq:3}) and the corresponding scores $\psi(\frac{|D_{gi}|}{h_{0}}) , 1\leq g \leq G, 1\leq i \leq I_g$ to conduct the test for joint comparison as described in Section \ref{submax}.
\subsection{The proposed method}
\label{groupm}
We are ready to introduce our novel scoring approach that accounts for the subgroup structures for the aforementioned trimming version of M-statistics. Instead of applying trimming for all the matched pairs simultaneously (i.e. scalar $h_0$ is applied for all $D_{gi}$), we employ this trimming procedure within each group $g$. Formally, denote $h_{g0} = \text{median}\{|D_{g1}|,|D_{g2}|,\ldots, |D_{gI_g}|\}$, and we have the following group M-statistic for each group $g, 1 \leq g \leq G$ that uses the subgroup structure:
\begin{equation}
\label{eq:4}
 T^{sub}_{g} =\sum_{i=1}^{I_g} sign(D_{gi})\psi(\frac{|D_{gi}|}{h_{g0}})h_{g0}.
\end{equation}
In particular, without the subgroup argument (i.e. $G=1$), it reduces to the M-statistic and M-score applied to all the matched pairs since there will be no difference if a constant (say $h_0$ here) is multiplied to all the observations in a permutation test. Note that when $G>1$, for each group $g$, we use the scalar $h_{g0}$ to do the trimming and retain the relative differences of the score sizes across different subgroups by multiplying the weight by the within-group median $h_{g0}$. The $\psi$ function is the same as in Section \ref{normalm} for all the numerical results in the article. The submax method based on the proposed equation (\ref{eq:4}) would have the same procedure as described previously in Section \ref{submax} but we will show that using the the proposed group M-statistic and the corresponding scores could result in improved performance in sensitivity analysis.

\begin{table}[hp]
\centering
\begin{tabular}{@{}cccccccc@{}}
\hline
\hline
$k$ & 1 & 2 & 3 & 4 & 5 & 6 & \\
\midrule
Subpopulation & All & Age $\leq 10$ & Age $> 10$ & Female & Male & Maximum \\ 
& $D_{\Gamma1}$ & $D_{\Gamma2}$ & $D_{\Gamma3}$ & $D_{\Gamma4}$ & $D_{\Gamma5}$  & $D_{\Gamma\text{max}}$  \\
\midrule
Sample-size & 1560 & 447 & 1113 & 766 & 794  \\
\midrule
& & & $\Gamma = 2.0$ & & & \\
Mean difference & \textbf{5.28} & \textbf{6.63} & -2.30 & \textbf{2.38} & \textbf{5.02}  & \textbf{6.63}  \\
M-statistic & 2.03 & \textbf{4.86} & -2.55 & 0.26 & \textbf{2.5}  & \textbf{4.86}  \\
Group M-statistic & \textbf{6.21} & \textbf{6.73} & -2.89 & \textbf{2.74} & \textbf{5.82} & \textbf{6.73}  \\
& & & $\Gamma = 2.7$ & & & \\
Mean difference & \textbf{2.52} & \textbf{4.74} & -5.00 & 0.48 & \textbf{3.02}  & \textbf{4.74}  \\
M-statistic & -2.10 & \textbf{2.31} & -5.92 & -2.63 & -0.44  & \textbf{2.31}  \\
Group M-statistic & \textbf{3.68} & \textbf{4.74} & -6.63 & 0.97 & \textbf{4.01}  & \textbf{4.74}  \\
& & & $\Gamma = 3.0$ & & & \\
Mean difference & 1.57 & \textbf{4.11} & -5.97 & -0.19 & \textbf{2.34}  & \textbf{4.11}  \\
M-statistic & -3.56 & 1.43 & -7.14 & -3.66 & -1.48  & 1.43  \\
Group M-statistic & \textbf{2.81} & \textbf{4.07} & -7.98 & 0.35 & \textbf{3.40}  & \textbf{4.07}  \\
& & & $\Gamma = 4.0$ & & & \\
Mean difference & -1.01 & \textbf{2.42} & -8.71 & -2.01 & 0.50 & \textbf{2.42}  \\
M-statistic & -7.60 & -0.96 & -10.55 & -6.52 & -4.34  & -0.96  \\
Group M-statistic & 0.48 & \textbf{2.29} & -11.78 & -1.33 & 1.76  & \textbf{2.29}  \\
& & & $\Gamma = 4.1$ & & & \\
Mean difference & -1.24 & \textbf{2.28}  & -8.95 & -2.17 & 0.35  & \textbf{2.28}  \\
M-statistic & -7.95 & -1.17 & -10.86 & -6.77 & -4.58  & -1.17  \\
Group M-statistic & 0.28 & 2.14 & -12.11 & -1.48 & 1.63  & 2.14  \\
\bottomrule
\end{tabular}
\caption{Sensitivity analysis by using submax method with three test statistics to the malaria control data. We are interested in two binary covaries, age and gender, here, so the submax method would return five test statistics and we reject the null hypothesis if the maximum of these five statistics is larger than the corresponding critical values. The critical values for using mean difference, M-statistic, and group M-statistic are 2.20, 2.20, 2.18 respectively. The test statistics larger than the critical values are in bold. $\Gamma = 2.7$ and $\Gamma = 4.1$ are selected as the largest sensitivity values, reported to one decimal place, for which the null hypothesis can still be rejected using the submax method with the M-statistic and the mean-difference statistic respectively.}
\label{tab:garki_apply}
\end{table}

\section{Application in the malaria example}
\label{garki}
As shown in Table \ref{tab:garki_apply}, we applied the proposed group M-statistic in section \ref{groupm} to the malaria control example. In this example, effect modification takes place in the subgroup of young people (with age less than $10$ years old), and the maximum statistic is always based on those $447$ pairs of young people. It is notable that if we use the raw data (i.e. mean difference statistic) as the score, we still have evidence to reject the null hypothesis of no treatment effect up to the sensitivity value at $\Gamma = 4.1$; however, if we use the M-statistic by trimming all the matched pairs together, then we have evidence up to the sensitivity value only at $\Gamma = 2.7$. In other words, if the researcher wants to use M-statistics to make a robust argument, the reported insensitivity to bias is substantially less than using the raw data. Intuitively, since all the treated-minus-control differences are scored together, those treated-minus-control differences with large and positive values from some subgroup where effect modification happens are affected most by the trimming in the M-statistic, resulting in the  unwanted smaller $\Gamma$. 
Our proposed group M-statistic in section \ref{groupm} employs the trimming process within each subgroup and Table \ref{tab:garki_apply} shows that we are still able to reject the null hypothesis of no treatment effect with a sensitivity value at $\Gamma = 4.0$, comparable to using the raw data. 

To connect with common practice in the application, we also conducted matched pair regression. Note that this part of analysis is of separate interest. A mixed effect model for matched pair regression was used and age and gender were included as covariates. We found that no detectable between-pair variability in baseline outcomes after accounting for age and sex, while the treated outcome is on average $12.7$ units lower than the control outcome.

\section{Simulation}
\label{sim}

In this section, we examine the power of sensitivity analysis in finite samples of the proposed method in extensive settings to show the superior performance compared with using raw data and the conventional practice of using M-statistics. We evaluate the power of sensitivity analysis as in other papers under the "favorable situation" that there is in fact no unmeasured confounding \citep{effect_mod,cluster}. However, if a researcher is in this favorable situation, it could not be told from the observed data. Then, a sensitivity analysis is still performed and perhaps the best we can hope to say is that the study is insensitive to certain unmeasured confounding quantified by sensitivity parameter $\hat{\Gamma}$. It is in this favorable situation that we know we want to reject the null hypothesis of no treatment effect and the power of sensitivity analysis is the probability that such hope can be realized. Now consider we have two binary covariates of interest and one of them is set to be the effect modifier. A favorable situation in the effect modification setting would be that we pre-specify the constant effect sizes $\tau_1$ and $\tau_2$ for different subgroups and sample the independent error terms $\epsilon_1$ and $\epsilon_2$ that might be different across subgroups (as we are sampling conditional on the covariate) but have a symmetric distribution centered at zero and i.i.d within a group since the treatment is assigned at random. 
Specifically, suppose we now have $I = 1000$ matched pairs with two binary covariates and $I_{g} = 250, 1\leq g \leq 4$ in the four groups formed by the two binary covariates. Effect modification only happens with the first covariate. Specifically, the first $500$ matched pairs have value $1$ of the first covariate and the second $500$ matched pairs have value $0$, and the effect of the treatment is larger in the first $500$ matched pairs.

\begin{figure}[p]
    \centering
    \includegraphics[width=\textwidth]{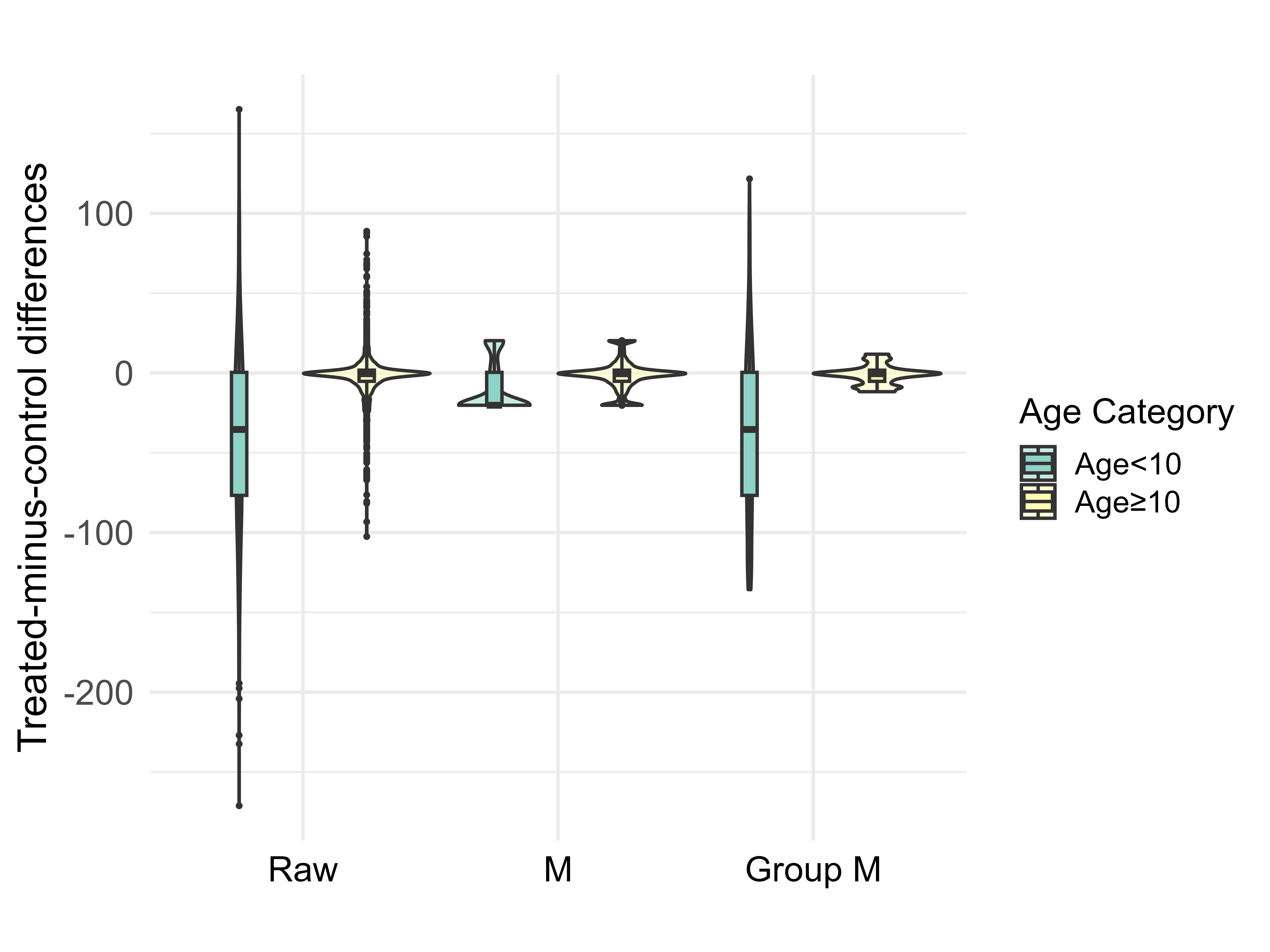}
    \caption{Treated-minus-control differences across different statistics for people with age less than 10 and larger than 10 in the malaria control data. For the M-statistic, we multiply back the parameter $h_0$ in the figure to make the comparisons on the same scale with the raw data and the group M-statistic. It is essentially the same when we do inference.
    }
    \label{fig:garki_stats}
\end{figure}

\begin{table}[hp]
\centering
\begin{tabular}{lccccc}
\hline
\hline
Sampling Situation & $\hat{\Gamma}$ & Mean difference & M-statistic & Group M-statistic \\
\hline
Situation (1)
& 1 & 1.000 & 1.000 & \textbf{1.000} \\
& 2 & 0.997 & 0.887 & \textbf{0.998}\\
& 3 & \textbf{0.151} & 0.002 & 0.133 (-0.018) \\
& 4 & 0.000 & 0.000 & \textbf{0.000} \\
\hline
Situation (2) 
 & 1 & 1.000 & 1.000 & \textbf{1.000} \\
 & 2 & 1.000 & 1.000 & \textbf{1.000} \\
 & 3 & 0.998 & 1.000 & \textbf{1.000} \\
 & 4 & 0.769 & 0.848 & \textbf{0.926} \\
 & 5 & 0.186 & 0.153 & \textbf{0.296} \\
\hline
Situation (3) 
 & 1 & 1.000 & 1.000 & \textbf{1.000} \\
 & 2 & 1.000 & 1.000 & \textbf{1.000} \\
 & 3 & 1.000 & 0.998 & \textbf{1.000} \\
 & 4 & \textbf{0.993} & 0.588 & 0.991 (-0.002) \\
 & 5 & \textbf{0.675} & 0.051 & 0.656 (-0.019) \\
\hline
Situation (4) 
 & 1 & 1.000 & 1.000 & \textbf{1.000} \\
 & 2 & 1.000 & 1.000 & \textbf{1.000} \\
 & 3 & 0.965 & 0.991 & \textbf{0.998} \\
 & 4 & 0.504 & 0.346 & \textbf{0.708} \\
 & 5 & 0.090 & 0.012 & \textbf{0.146} \\
\hline
Situation (5) 
 & 1 & 1.000 & 1.000 & \textbf{1.000} \\
 & 2 & 1.000 & 1.000 & \textbf{1.000} \\
 & 3 & 0.939 & \textbf{0.998} & 0.997 (-0.001) \\
 & 4 & 0.445 & \textbf{0.707} & 0.673 (-0.043) \\
 & 5 & 0.070 & \textbf{0.139} & 0.127 (-0.012) \\
\bottomrule
\end{tabular}
\caption{Simulated power of sensitivity analysis under different sampling situations for $I=1000$ matched pairs. Each sampling situation is repeated 10,000 times and the power is evaluated by the proportion of sharp nulls that were rejected at significance level 0.05 in each case. There are two binary covariates of interest and effect modification happens for the first $500$ matched pairs. The details have been described in the main text. The power of group M-statistics in each sampling situation is in bold if it achieves the highest power (including tied results); otherwise, we show in parenthesis the difference between the power of group M-statistics and the statistics that achieves the highest power and the corresponding power of that test statistic is in bold.}
\label{tab:simu}
\end{table}
The power of sensitivity analysis in finite samples is evaluated by repeating each sampling situation for $10,000$ times and calculating the proportion of sharp nulls that were rejected for each test statistic at significance level 0.05. Inspired by the distribution of treated-minus-control differences of young children and older people in the malaria control data in Figure \ref{fig:garki_stats}, we further consider the following sampling situations of treated-minus-control difference for $I=1000$ independent matched pairs:

(1) $D_i = 5+10\epsilon_{1i}, 1\leq i \leq 500$, where $\epsilon_{1i}$ is standard normal distribution; $D_i = 0.5+\epsilon_{2i}, 501\leq i \leq 1000,$ where $\epsilon_{2i}$ is t distribution with degree of freedom 3.

(2) $D_i = 5+5\epsilon_{i}, 1\leq i \leq 500$; $D_i = 0.5+0.5\epsilon_{i}, 501\leq i \leq 1000,$ where $\epsilon_{i}$ is t distribution with degree of freedom 3.

(3) $D_i = 4+5\epsilon_{i}, 1\leq i \leq 500$; $D_i = 0.2+\epsilon_{i}, 501\leq i \leq 1000$, where  $\epsilon_{i}$ is standard normal distribution.

(4) $D_i = 5+5\epsilon_{1i}, 1\leq i \leq 500$, where $\epsilon_{1i}$ is t distribution with degree of freedom 3; $D_i = 0.2+0.5\epsilon_{2i}, 501\leq i \leq 1000$, where $\epsilon_{2i}$ is standard normal distribution.

(5) $D_i = 1+\epsilon_{i}, 1\leq i \leq 500$ and $D_i = 0.5+\epsilon_{i}, 501\leq i \leq 1000,$ where $\epsilon_{i}$ is t distribution with degree of freedom 3.

In cases (1) and (4) the error term comes from two distinct families of distribution (normal and t distribution); in cases (1), (2), (3), and (4), the first $500$ matched pairs have substantial treatment effects accompanied with large variability while the rest of the $500$ matched pairs have smaller treatment effects and smaller variability in the error terms and it is in these cases that trimming all the matched pairs together would potentially confuse effect modification with outliers and result in smaller power given a sensitivity parameter $\hat{\Gamma}$; in case (5), we have moderate effect modification with the same t-distribution for all the $1000$ matched pairs and similar performance of using M-statistic and the proposed method is observed.

We note that the proposed group M-statistic performs well in all the settings as shown in Table \ref{tab:simu}. In particular, it achieves higher power given a sensitivity parameter $\hat{\Gamma}$ in sampling situation (2) and (4) where the error term follows a t-distribution in the subgroup with much larger treatment effect. If other statistics have higher power, then the differences are almost negligible whether it is achieved by using raw data as in situation (3) or using M-statistics in a conventional manner as in situation (5). Overall, the simulation results show that our group M-statistic that does subgroup-aware scoring works well across different settings.

\section{Design sensitivity}
In this section, we derive the large sample power and compute the design sensitivity of the proposed method to justify the observed improvements of our proposed method.
\subsection{Large sample power calculations and design sensitivity}
In a favorable situation where there is a treatment effect and no unmeasured confounding, we would like to evaluate the power of a sensitivity analysis. As in \cite{submax}, under an alternative hypothesis, we have $T_g$ asymptotically Normal with mean $\mu_g^*$ and variance $\nu_g^*$ for each $1\leq g \leq G$ when $\min(I_g) \to \infty$. Thus, we have the asymptotic distribution of $S_k = \sum_{g=1}^G c_{kg}T_g$. Denote $\theta_k^*=\sum_{g=1}^G c_{kg}\mu_g^*$ and ${\sigma^*_k}^2$ the $k^{th}$ diagonal element of $\mathbf{\Sigma}^* = \mathbf{C V}^* \mathbf{C}^T$, where $\mathbf{V}^*=\text{diag}\{\nu_1^*,\ldots,\nu_G^*\}$. Thus, we know that $\theta_k^*$ and $\sigma_k^*$ are the mean and standard deviation of $S_k$ respectively. Finally, write $\mathbf{\rho}^*$ for its correlation matrix with the element at $i^{th}$ row and $j^{th}$ column derived from dividing $(\boldsymbol{\Sigma}^*)_{ij}$ by $\sigma^*_i$ and $\sigma^*_j$. Given a sensitivity parameter $\Gamma$, the power based on our test statistics for the $k^{th}$ component test $D_{\Gamma k}$ and the joint comparison $D_{\Gamma \text{max}}$ can be calculated under the alternative hypothesis specified by some model generating the treated-minus-control differences. Specifically, as in \citet{submax}, the power of the joint comparison is the following probability
\begin{equation}
\label{eq:5}
    \begin{aligned}  \operatorname{Pr}\left( D_{\Gamma \text{max}} \geq \kappa_{\Gamma,\alpha} \right) &= 1-\operatorname{Pr}\left( D_{\Gamma \text{max}} < \kappa_{\Gamma,\alpha} \right)\\& = 1 - \operatorname{Pr}\left(\frac{S_k-\theta_{\Gamma k}}{\sigma_{\Gamma k}} < \kappa_{\Gamma, \alpha}, k=1, \ldots, K\right) \\ & =1 - \operatorname{Pr}\left(\frac{S_k-\theta_k^*}{\sigma_k^*}<\frac{\theta_{\Gamma k}-\theta_k^*+\kappa_{\Gamma, \alpha} \sigma_{\Gamma k}}{\sigma_k^*}, k=1, \ldots, K\right) .\end{aligned}
\end{equation}
This could be calculated by $\boldsymbol{N}_K(\boldsymbol{0},\boldsymbol{\rho}^*)$. Similarly, for the $k^{th}$ component test using $D_{\Gamma k}$ only (suppose $k$ is fixed), the power is
\begin{equation}
\label{eq:6}
\begin{aligned}  \operatorname{Pr}\left( D_{\Gamma k} \geq \Phi^{-1}(1-\alpha) \right) &=  \operatorname{Pr}\left( \frac{S_k-\theta_{\Gamma k}}{\sigma_{\Gamma k}}\geq \Phi^{-1}(1-\alpha)  \right)\\ & =  1-\operatorname{Pr}\left\{\frac{S_k-\theta_k^*}{\sigma_k^*}<\frac{\theta_{\Gamma k}-\theta_k^*+\Phi^{-1}(1-\alpha) \sigma_{\Gamma k}}{\sigma_k^*}\right\},\end{aligned}
\end{equation}
which can be calculated by the standard normal distribution.

Under the same assumptions in a favorable situation, there is typically a value $\widetilde{\Gamma}$ called the design sensitivity such that the power of the test (equation (\ref{eq:5}) and equation (\ref{eq:6})) tends to $1$ if $\Gamma > \widetilde{\Gamma}$, and tends to zero if $\Gamma<\widetilde{\Gamma}$ when $\text{min}(I_g) \to \infty$. As shown in \citet{submax} and in our equation (\ref{eq:5}), for the submax method using $D_{\Gamma \text{max}}$, if $\Gamma < \widetilde{\Gamma}_k$ for some $k, 1\leq k \leq K$, then $\operatorname{Pr}\left(D_{\Gamma k}\geq\kappa_{\Gamma,\alpha} \right)$ tends to 1 for arbitrary $\kappa_{\Gamma, \alpha}$, and thus $\operatorname{Pr}\left(D_{\Gamma \text{max}}\geq\kappa_{\Gamma,\alpha} \right)$ tends to 1 as $D_{\Gamma \text{max}}$ is the maximum of $\{D_{\Gamma 1},\dots,D_{\Gamma K}\}$. Conversely, if $\Gamma \geq \widetilde{\Gamma}_k$ for all $1\leq k \leq K$, then the power by using $D_{\Gamma \text{max}}$ tends to zero. Thus, the design sensitivity of $D_{\Gamma\text{max}}$, denoted as $\widetilde{\Gamma}_{\text{max}}$, is the largest design sensitivity of all $K$ component tests, i.e. $\widetilde{\Gamma}_{\text{max}} = \text{max}\{{\widetilde{\Gamma}}_1,\dots,{\widetilde{\Gamma}}_K\}$. Consequently, calculating the design sensitivity of the submax method reduces to calculating the design sensitivity for each component test. The design sensitivity $\widetilde{\Gamma}_k$ for $D_{\Gamma k}$ is the value of $\Gamma$ that solves $1 = S_k/{\theta_{\Gamma k}}, \min(I_g)\to \infty$ under an alternative hypothesis \citep{submax}.

Some mild regularity conditions for the existence of the design sensitivity by using M-statistics are detailed in \citet{impact_multiple_m}, which essentially covers the commonly used Huber M-statistics and the mean difference statistics. Simpler formulas for the design sensitivity of M-statistics are available in the case of matched pairs as in \citet{impact_multiple_m}, and the result applies to our proposed group M-statistics. In the next section, we explicitly derive the formula for design sensitivity of using the submax method based on group M-statistics in a simple model. We will also numerically evaluate the design sensitivity of different statistics for all the data-generating processes in the previous simulations. 

\subsection{Design sensitivity of proposed method in a simple case}
To calculate the design sensitivity, we specify the model and the data generating process of the treatment-minus-control differences. Consider a simple model similar to \citet{submax} where we have $L=2$ binary covariates and a balanced sample size of each interaction group $g, 1\leq g \leq G = 2^L = 4$, say $I_g = I/2^L = I/4$. Effect modification happens in the subgroup indicated by the first covariate and is independent of another covariate. This simple model is consistent with the settings in our simulations in Section \ref{sim}. Denote $\widetilde{\Gamma}_k$  as the design sensitivity for the $k^{th}$ component test, and it is the value that solves $1= S_k/\theta_{\Gamma k}$ when $\min (I_g)\to\infty$, that is 
\begin{equation}
\label{eq:7}
\frac{\sum_{g=1}^{G}c_{gk}T_g}{\sum_{g=1}^{G}c_{gk}I_g} = \frac{{\sum_{g=1}^{G}c_{gk}\mu_{\Gamma g}}}{\sum_{g=1}^{G}c_{gk}I_g}, \min(I_g)\to\infty.
\end{equation}
Note that equation (\ref{eq:7}) is very similar to the form in Section 3.5 of \citet{submax}, but we use $T_g$ instead of directly using $\mu_g^*$ in our notation, and normalize the quantity by $\sum_{g=1}^{G}c_{gk}I_g$, as this will help when we plug in the group M-statistics in the following discussions and is connected to the form in \citet{impact_multiple_m}.
For matched pairs, denote the distribution function of the treated-minus-control difference as $F_g$ in each group $g$. As $\min(I_g) \to \infty$, the scalar $h_{g0}$ of the group M statistics in equation (\ref{eq:4}) converges in probability to the $\sigma_g$ that solves
\begin{equation}
\label{eq:8}
\frac{1}{2}= \int_{-\sigma_g}^{\sigma_g} d F_g(y).
\end{equation}
Following the results for the matched pairs by using M-statistics in \citet{impact_multiple_m} and equation (\ref{eq:7}), we have 
\begin{equation}
\label{eq:9}
\frac{1}{\sum_{g=1}^Gc_{gk}}\left( \mathbb{E}\left[ \sum_{g=1}^G c_{gk}\psi\left(\frac{D_g}{\sigma_g}\right)\sigma_g \right]  \right) = \frac{1}{\sum_{g=1}^Gc_{gk}}\frac{\Gamma-1}{\Gamma+1}\left( \mathbb{E}\left[ \sum_{g=1}^G c_{gk} \psi\left(\frac{|D_g|}{\sigma_g}\right) \sigma_g\right]  \right) ,
\end{equation}
where we used the assumption of balanced sample by $\sum_{g=1}^Gc_{gk}I_g = \left(\sum_{g=1}^Gc_{gk}\right)I_1$. Solving (\ref{eq:9}), we have the following formula for the design sensitivity of $k^{th}$ comparison by using group m-statistics:

\begin{equation}
\label{eq:10}
\widetilde{\Gamma}_k =\frac{ \sum_{g=1}^G c_{gk} \sigma_g \int_{0}^{\infty}\psi\left( \frac{|y|}{\sigma_{g}} \right)dF_g(y)}{\sum_{g=1}^G c_{gk} \sigma_g \int_{-\infty}^{0}\psi\left( \frac{|y|}{\sigma_{g}} \right)dF_g(y)}
\end{equation}
\begin{table}[t]
\centering
\begin{tabular}{lccccc}
\hline
\hline
Sampling Situation  & Mean difference & M-statistic & Group M-statistic \\
\hline
Situation (1)
 & \textbf{3.528} & 2.917 & 3.467 \\
\hline
Situation (2) 
 & 5.587 & 5.303 & \textbf{5.955} \\
\hline
Situation (3) 
 & \textbf{7.655} & 5.486 & 7.573 \\
\hline
Situation (4) 
 & 5.587 & 4.997 & \textbf{5.955} \\
\hline
Situation (5) 
 & 5.587 & \textbf{5.993} & 5.955 \\
\bottomrule
\end{tabular}
\caption{Design sensitivities under different sampling situations across mean difference, m-statistic and the proposed group m-statistic when $I_g\to \infty, 1\leq g \leq G=4$. The details of the sampling situations could be found in the main text. Briefly, for each interaction group, we have balanced number of matched pairs where the distribution of the treatment-minus-control differences could be different. We have two binary covariates of interest and effect modication happens for the first $500$ matched pairs.}
\label{tab:designsen}
\end{table}


To illustrate, we use sampling situation 4 in our simulations as an example. We denote $c_1$ and $c_2$ as two binary covariates of interest. Suppose that $c_1 = 1$ indicates the individuals in interaction group $g=1$ and $g=2$ with treated-minus-control differences $D = 5+5\epsilon_1$ where $\epsilon_1$ has a t distribution with degree of freedom 3. Similarly, we set $c_1=0$ as individuals in interaction group $g=3$ and $g=4$ with treated-minus-control differences as $D = 0.2+0.5\epsilon_2$ where $\epsilon_2$ has a standard normal distribution. Set $c_2 = 1$ for interaction groups $g=1,3$, and $c_2=0$ otherwise. Suppose only matched pairs with $c_1=1$ are included in the second component test $k =2$, and thus we have $c_{12} = c_{22} =1, c_{32} = c_{42} =0$ while $F_g, g=1,2$ are the distribution functions of treated-minus-control in interaction group $g=1$ and $g=2$, which are the same. The design sensitivity of this test is $\widetilde{\Gamma}_2 = 5.955$. Similarly, the test based on $c_1=0$, say the third component test $k=3$, has design sensitivity $\widetilde{\Gamma}_3 = 2.690$. All the matched pairs are included in the test for $k=1$, while $k=4,5$ indicates the tests based on subgroups formed by $c_2$. From equation (\ref{eq:10}), it is easy to see that the design sensitivities are the same across the rest of tests, say $k = 1,4,5$, and this design sensitivity is between $\widetilde{\Gamma}_2$ and $\widetilde{\Gamma}_3$ because they reduce to the same form as 

$$\left(\sum_{g=2}^3 \sigma_g \int_{0}^{\infty}\psi\left( \frac{|y|}{\sigma_{g}} \right)dF_g(y)\right)/\left(\sum_{g=2}^3 \sigma_g \int_{-\infty}^{0}\psi\left( \frac{|y|}{\sigma_{g}} \right)dF_g(y)\right),$$
which is essentially the sum of the nominator of $\widetilde{\Gamma}_2$ and $\widetilde{\Gamma}_3$ over the sum of the denominator of $\widetilde{\Gamma}_2$ and $\widetilde{\Gamma}_3$ in our model. Thus, the design sensitivity of submax method by using group M-statistics in this case is $\widetilde{\Gamma}_{\max} = \widetilde{\Gamma}_2 = 5.955$. Design sensitivities of group M-statistics and other statistics in all the different sampling situations mentioned in Section \ref{sim} are provided in Table \ref{tab:designsen}. In the next section, we provide more discussions.

\subsection{Design sensitivities of different statistics by using submax method}
The design sensitivity of using conventional M-statistics is computed by using a fixed scalar $\sigma_0$ for all the component tests in equation (\ref{eq:10}). $\sigma_0$ is derived from the entire population which has a mixture distribution with equal probability from $g_1$ and $g_2$ in our specified model above. Thus, as is the case in sampling situation 4, the mixture distribution will underestimate the scalar parameter for the second test where effect modication plays a role in this subgroup. Meanwhile, the mean difference test statistic applies no trimming and uses the original treated-minus-control in the test. Table \ref{tab:designsen} shows the complete results of the design sensitivities across different statistics in all the sampling situations described in Section \ref{sim}. Consistent with Table \ref{tab:simu}, the design sensitivity of using the proposed group M-statistic is larger than the conventional M-statistic in almost all the situations and remains a robust choice across all the sampling situations in the sense that it has superior design sensitivity or has negligible difference with the superior test statistic.

The scalar parameter is pivotal in combining the submax method and M-statistics. In Table \ref{tab:designsen}, we note that sampling situation (2), (4), and (5) have the same design sensitivities for mean difference test statistic and the proposed group M-statistic. This is presumably because the design sensitivity comes from the same component test for the subgroup with a larger treatment effect. However, using M-statistics would yield different design sensitivities because now the distributions of the treated-minus-control differences distribution of the entire population are not the same due to the different mixture components of the treated-minus-control differences from the rest of sample, and thus the scalar shared by all the component tests across different sampling situations results in the difference in the design sensitivities. Moreover, the treatment effect is much lower in another subgroup, resulting in a much smaller scalar parameter of $\sigma_0$ calculated from the entire population compared with the scalar parameter calculated from the subgroup with a larger treatment effect. Specifically, if there is indeed a large treatment effect in one subgroup, then the underestimated scalar parameter would cause over-trimming, and the test statistic for this group with a larger treatment effect will be similar to a sign test and the information of effect sizes will be less relevant. 
In sampling situation (4), the effect size of the subgroup with smaller treatment effect is even smaller than that of sampling situation (2), while the error term has a lighter tail compared with situation (2). As a result, the design sensitivity of using conventional M-statistic is affected more by underestimating the scalar parameter, and thus is much smaller compared to situation (2), while in both cases the design sensitivities by using M-statistic are much lower than using mean difference statistic or the proposed group M-statistic. Conversely, in sampling situation (5) the scalar parameter estimated from the entire population has minor difference with the scalar parameter estimated from each subgroup as the two groups have similar effect sizes. 

Now that we are aware that different performances of these test statistics in terms of design sensitivity are caused by the trimming parameter, we end our discussion by providing more explanations on why underestimating the trimming parameter by using conventional M-statistic would necessarily yield a lower (as opposed to a superior) design sensitivity in some cases. This line of discussion with respect to the power of a sensitivity analysis aligns with the arguments in \citet{designsens_efficiency}. Heuristically, for a treatment effect to be distinguishable from the unmeasured confounding modeled by parameter $\Gamma$, the value of treated-minus-control difference should have at least $\Gamma/(1+\Gamma)$ probability to be positive to favor the rejection of the test. This can also be read from equation (\ref{eq:9}), as we hope that the score $\psi\left(\frac{D_g}{\sigma_g}\right)$  assigns higher weights for those values of $D_g$ with higher probability being positive in order to allow for a larger design sensitivity. Thus, the score function should attach higher weights for those values of treated-minus-control differences with more probability to be positive than to be negative for larger design sensitivity or larger power in the finite sample. Meanwhile, such values depend on the alternative distribution of treated-minus-control differences and \citet{design_sen} discussed several forms of distributions. In our settings, if the scalar parameter in the trimming process is underestimated, then most values would be assigned similar weights and, in particular, substantial weights are attached to many small values of treated-minus-control differences (and those small values typically have less chances to be positive compared with other values). This will reduce the power of sensitivity analysis as pointed in \citet{designsens_efficiency} and is consistent in our observations in sampling situations (1), (2), (3), (4). Moreover, in sampling situations (2), (4), and (5), we observe higher design sensitivities by using group M-statistic than mean difference statistic as the influence of the extreme values in the heavy-tail distribution has been capped by trimming. In practice, using M-statistics with trimming effect is also robust to outliers caused my measurement errors. Again, all these results revealed that by using within group trimming, the group M-statistic is a robust choice with superior design sensitivity for conducting sensitivity analysis involving subgroup comparisons.

\section{Discussion}
In this article, we introduced a novel scoring strategy that accounts for the subgroup structures when using M-statistics. By scoring the matched pairs within each subgroup, the proposed method works well with the malaria control data and reports a larger sensitivity value when rejecting the null hypothesis than using the conventional M-statistics while still being robust to the outlier. We demonstrated that the proposed method is suitable in various settings and enjoys superior or non-inferior performance in terms of the power in finite samples and the design sensitivity compared to using raw data (that lacks robustness) and using conventional M-statistics. The key difference is that using group M-statistics correctly estimates the trimming parameter within each subgroup and thus assigns more appropriate weights \citep{designsens_efficiency} than conventional M-statistics in certain situations.

\begin{figure}[p]
    \centering
    \includegraphics[width=\textwidth]{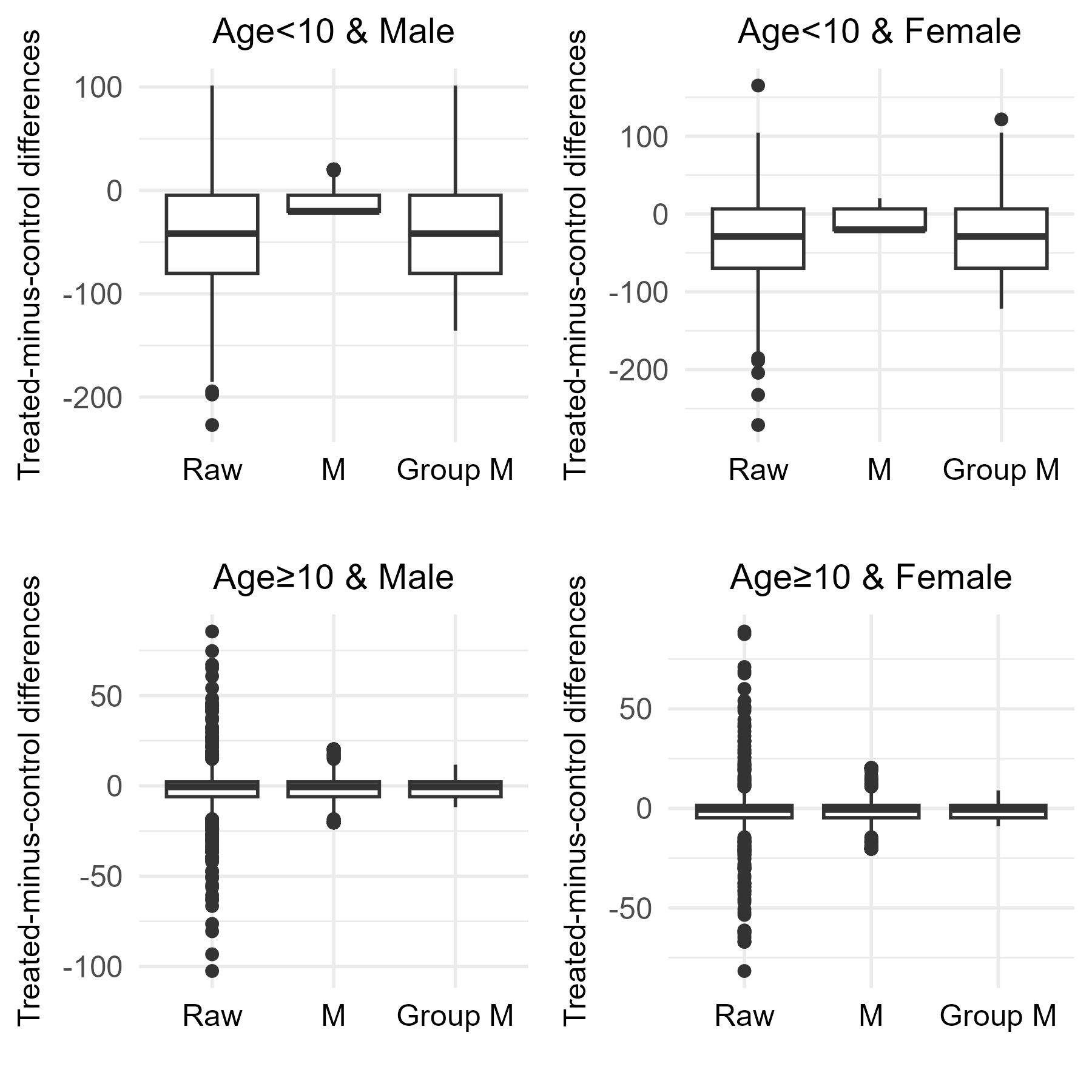}
    \caption{Treated-minus-control differences across different statistics for people in four non-overlapped subgroup defined by age and gender. For the M-statistic, we multiply back the parameter $h_0$ in the figure to make the comparisons on the same scale with the raw data and the group M-statistic.}
    \label{fig:trim}
\end{figure}

Importantly, the practice of scoring all the matched pairs together when using conventional M-statistics would potentially confuse the outliers with effect modification by wrongly estimating the trimming parameter. This practice does not distinguish whether the scores at the tail of the treated-minus-control differences come from the subgroup with a large treatment effect (e.g. people with age less than ten in the malaria control example) or from the randomness of long tail distribution of the error term. Figure \ref{fig:garki_stats} shows the resulting distribution of treated-minus-control scores for different age groups given by three test statistics. We note that the contribution of pairs from people with age less than ten was trimmed away when using conventional M-statistics, which made the information of effect sizes less relevant in the test and resulted in inferior performance in sensitivity analysis. By contrast, the proposed group M-statistics preserved pairs with large treatment effect and Figure \ref{fig:trim} further shows that the outliers are properly trimmed in each interaction group by our proposed method.

Moreover, it is worth mentioning that the meaning of the test and the sensitivity analysis conducted by using submax method is to view the population as a whole and test the global null hypothesis of no treatment effect \citep{submax}. If we would like to make further inferences for different subgroups, then it is required that the scores used in a given component test should only depend on the matched pairs within these subgroups involved in the test, and our proposed method could be naturally used in such scenario while the conventional M-statistics would not be feasible (see further discussion in section 4 of \citet{submax}).
Meanwhile, as is the case in the malaria control data, the error term of different subgroups could be different since now we condition on a covariate. The interplay between the error distributions and the effect sizes across different subgroups affects the trimming parameter, and thus results in different scores (or weights) \citep{designsens_efficiency} attached to the values of treated-minus-control difference. They together explain how different statistics in sensitivity analysis behave in the setting of effect modification \citep{submax,effect_mod}. 
We used the example of Huber's M-statistics to illustrate the proposed subgroup-aware scoring approach and it is straightforward to extend this idea to other scoring procedures when
we would like to make the scoring function dependent on the covariates. In general, if we want to leverage the effect modification or conduct the subgroup comparisons, it is critical to be aware of subgroup structure throughout the analysis procedure.



\acks{The authors thank Eric Sun for helpful discussions and preparation of an initial version of the figures.}


\newpage

\appendix
\section*{Appendix A.}
\label{rpackage}
An \verb|R| package \verb|groupmscorev| for implementation is available at \href{https://github.com/yjf326/groupmscorev}{Github}.

\section*{Appendix B.}
\label{supp_figures}
Two supplementary figures are provided for visualizing the consequent distribution of treated-minus-control differences by using three test statistics on malaria control data.


\vskip 0.2in
\bibliography{sample}

\end{document}